\documentclass[12pt]{article}
\usepackage{graphicx}
\def\g{{\cal G}^{++}}
\def\G{{\cal G}^{+++}}
\def\cm{{\cal M}}
\begin{document}
\thispagestyle{empty}
\setcounter{page}{0}
\renewcommand{\theequation}{\thesection.\arabic{equation}}

{\hfill{\tt hep-th/0412184}}

{\hfill{ULB-TH/04-27}}

\vspace{1cm}

\begin{center} {\bf \large  From  very-extended to overextended gravity
and M-theories}

\vspace{.5cm}

Fran\c cois Englert${}^{a,c}$, Marc Henneaux${}^{b,c}$ and Laurent
Houart${}^{b,c}$\footnote{Research Associate F.N.R.S.}

\footnotesize
\vspace{.5 cm}

${}^a${\em Service de Physique Th\'eorique, Universit\'e Libre de
Bruxelles,
\\ Campus Plaine, C.P.225\\Boulevard du Triomphe, B-1050 Bruxelles,
Belgium}

\vspace{.2cm}

${}^b${\em Service de Physique Th\'eorique et Math\'ematique,
Universit\'e Libre de Bruxelles,
\\ Campus Plaine C.P. 231\\ Boulevard du Triomphe, B-1050 Bruxelles,
Belgium}

\vspace{.2cm}

${}^c${\em The International Solvay Institutes, \\Campus Plaine C.P.
231\\ Boulevard du Triomphe, B-1050 Bruxelles, Belgium}\\ {\tt
fenglert, henneaux, lhouart@ulb.ac.be}

\end{center}

\vspace {1.5cm}

\centerline{\bf Abstract}

\noindent
The formulation of gravity and M-theories as very-extended
Kac-Moody invariant theories  encompasses, for each very-extended algebra
G+++,  two distinct actions invariant under the  overextended Kac-Moody
subalgebra G++. The first  carries a Euclidean signature and is the
generalisation to G++  of the  E10-invariant action proposed in the
context of M-theory and cosmological billiards. The second action carries
various Lorentzian signatures  revealed through various equivalent formulations
related by Weyl transformations of fields. It admits   exact
solutions, identical to those of the maximally oxidised field theories and
of their exotic counterparts, which  describe intersecting extremal branes
smeared in all directions but one. The Weyl transformations of G++ relates
these solutions by conventional and exotic dualities. 
These exact solutions, common to the Kac-Moody theories and to
space-time covariant theories, provide a laboratory for analysing the
significance of the infinite set of fields appearing in the Kac-Moody
formulations.

\newpage
\baselineskip18pt

\setcounter{equation}{0}
\addtocounter{footnote}{-1}

\section{Introduction}

A maximally oxidised theory  is a Lagrangian theory of  gravity coupled to forms
and dilatons defined in the highest possible space-time dimension $D$ which
upon dimensional reduction  to three dimensions exhibits the symmetry of
a simple Lie group ${\cal G}$ realised on a coset space
${\cal G}/{\cal H}$. Here
$\cal H$ is the maximal compact subgroup of $\cal G$. The maximally
oxidised actions  have been constructed for all   $\cal
G$ \cite{cjlp}.  They  comprise in particular pure gravity in $D$
dimensions and the low energy effective actions of the bosonic string and
of  M-theory.
\begin{figure}[h]
   \centering
   \includegraphics[width=12cm]{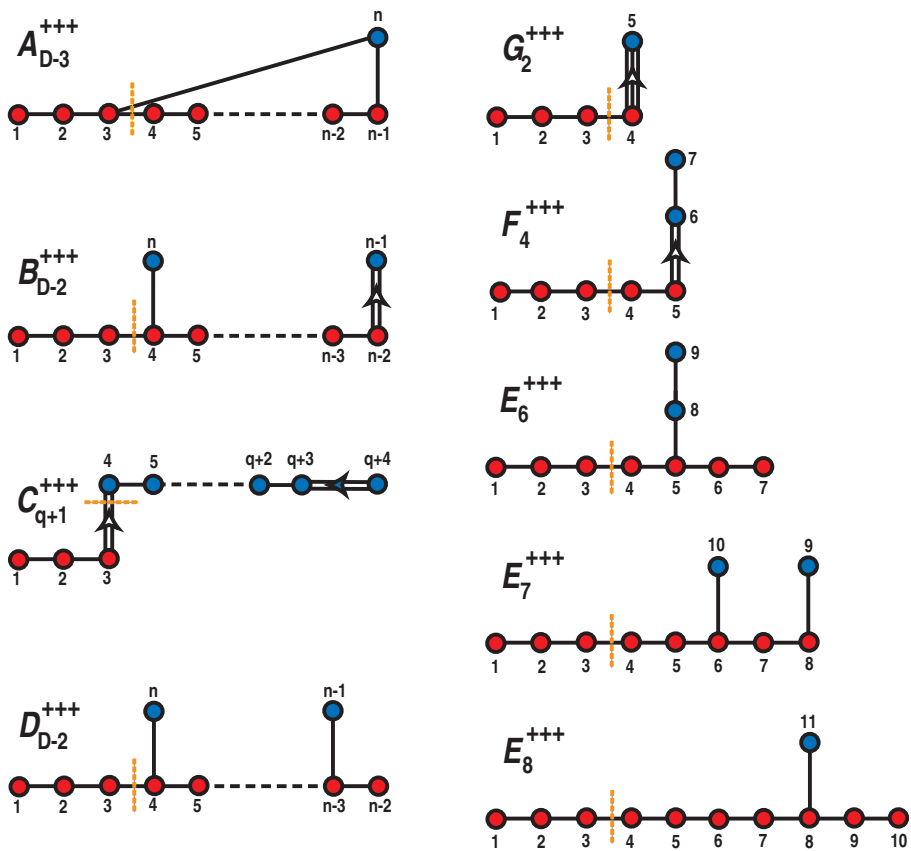}
 \caption { \small The
nodes labelled 1,2,3 define the Kac-Moody extensions of the  Lie
algebras. The horizontal line
starting at 1 defines the `gravity line', which is the
Dynkin diagram  of a
$A_{D-1}$ subalgebra. Note that for some Kac-Moody algebras the choice of a
gravity line is
 not unique, in which case we take here the one shown in the figure.}
   \label{first}
\end{figure}

It
has been conjectured that these theories, or some extensions of them,
possess the much larger very-extended Kac-Moody symmetry
$\G$. $\G$ algebras are defined by the Dynkin diagrams depicted
in Fig.1,  obtained from those of $\cal G$ by adding three
nodes~\cite{ogw}. One first adds the affine node, labelled 3 in the
figure, then a second node, 2,  connected to it by a single line and
defining the overextended ${\cal G}^{++}$ algebra,   then  a third
one, 1, connected  by a single line to the overextended node. Such
$\G$ symmetries were first conjectured in the aforementioned
particular cases
\cite{west01,lw} and the extension to all
$\G$ was proposed in
\cite{ehtw}. In a different development, the study of the properties
of cosmological solutions in the vicinity of a space-like singularity,
known as cosmological billiards
\cite{damourhn00},  revealed an  overextended symmetry ${\cal G}^{++}$
for all maximally oxidised theories
\cite{damourh00, damourbhs02}.

To explore the possible fundamental significance of these huge
symmetries two Lagrangian formulations \cite{damourhn02, eh}  {\it
explicitly} invariant under such infinite-dimensional Kac-Moody
algebras  have been proposed.

First  \cite{damourhn02} an
$E_{10}\equiv E_8^{++}$-invariant action was constructed in the
context of M-theory as a reparametrisation invariant $\sigma$-model of
fields depending on one parameter $t$ and living on the coset space
$E_8^{++}/K_8^{++}$. Here $K_8^{++}$ is the subalgebra of $E_8^{++}$
invariant  under the Chevalley involution. The action is built in a
recursive way  by a level expansion of $E_8^{++}$ with respect to its 
subalgebra $A_9$ whose Dynkin diagram is the `gravity line' defined in Fig. 1,
with the node 1 deleted.  The level of an irreducible representation of
$A_9$ occurring in the decomposition of the adjoint  representation of
$E_8^{++}$ counts the number of  times the simple root  not
pertaining to the gravity line appears in
the decomposition.  The $\sigma$-model, limited to the
real roots up to level 3, reveals  a perfect  match with the bosonic
equations of motion of 11-dimensional supergravity in the vicinity of
the  space-like singularity of the $E_8^{++}$ cosmological billiards,
where the fields depend only on time\footnote{See also \cite{damourn04} for an
analysis in a different formulation, and
\cite{nike} for a discussion of $E_8^{++}$ adapted to $IIB$
supergravity.}. In the dictionary relating the
$\sigma$-model to supergravity, the  parameter $t$
 is identified with time.  It was conjectured that   space derivatives
are hidden  in some higher level fields of the $\sigma$-model.
 This approach is straightforwardly generalised, as seen
below, to   actions $S_{{\cal G}_C^{++}}$ related to cosmological
billiards and invariant under the overextended $\g$. Such
$\g$-invariant formulation  singles out naturally a time coordinate.

The second  approach \cite{eh}  yields  a $\G$-invariant action $S_{\G}$
which puts  all the space-time coordinates on the same footing. This
is achieved by   a reparametrisation invariant $\sigma$-model with
fields depending on a parameter $\xi$ spanning a world-line a priori
unrelated to space-time. One uses a level decomposition of $\G$ with
respect to the subalgebra $A_{D-1}$ of its gravity line\footnote{ Level expansions of very-extended
algebras in terms of the subalgebra $A_{D-1}$ have been
considered in
\cite{west02, nifi, weke}.}  and
$D$ is identified to  the space-time dimension. The $\xi$-dependent
fields live in a coset space $\G /K^{+++}$ where the subalgebra
$K^{+++}$ is invariant under a temporal involution
\cite{eh} which preserve the Lorentz algebra $SO(1,D-1)$ and ensures
that the action $S_{\G}$ is Lorentz invariant
at each level.

In this paper, we analyse the $\g$ content of the
$\G$-invariant action $S_{\G}$. We find that the $\g$-invariant action
$S_{{\cal G}_C^{++}}$  is obtained by putting to zero in $S_{\G}$,
consistently with all its equations of motion, all the fields
which do not appear in $S_{{\cal G}_C^{++}}$. This consistent
truncation of the $\G$ theory implies that all the solutions of the equations of
motion of $S_{{\cal G}_C^{++}}$ are also solutions of the
equations of motion of $S_{\G}$. We then find that $S_{\G}$ contains another
 action $S_{{\cal G}_B^{++}}$ obtained by performing the
consistent truncation {\em after}  conjugation by a Weyl reflection in
$\G$. The action $S_{{\cal G}_B^{++}}$ is also invariant under a $\g$
algebra but is nonequivalent to $S_{{\cal
G}_C^{++}}$ and singles out naturally a space direction. Through Weyl
transformations in $\g$ expressed in terms of fields,
$S_{{\cal G}_B^{++}}$ can be formulated
 in various equivalent forms endowed with
various Lorentzian signatures $(-\dots -, +\dots +)$,  a consequence of
the non-commutativity of the temporal involution with Weyl reflections
\cite{keu1,keu2}. Exact solutions of $S_{{\cal G}_B^{++}}$ are obtained in
accordance with references \cite{eh,eh2}.  These solutions   are   {\em
identical} to those of the covariant Einstein and field equations describing,
depending on the signature, conventional or exotic intersecting extremal branes
smeared in all directions but one
\cite{aeh, ah}. They  transform into each other by
`duality' Weyl transformations. As all these brane solutions of
$S_{{\cal G}_B^{++}}$  are also solutions of the
$\G$-invariant action $S_{\G}$, the latter necessarily contains solutions
of exotic counterparts of the original maximally oxidised theories. In
particular the $E_{11}\equiv E_8^{+++}$-invariant action contains
in addition to the M-theory solutions, the exotic branes of the
related M' and M* theories \cite{hull1,hull2,ah, exop}.
Of course $S_{\G}$ can, as $S_{{\cal
G}_B^{++}}$, be formulated with different signatures related
by field transformations.

The fact that $S_{{\cal G}_B^{++}}$, and hence $S_{\G}$, contains
{\em exact} intersecting extremal  brane solutions of space-time
covariant theories compactified to all dimensions but one provides a
laboratory to analyse the significance of at least some subset of the
infinite many fields appearing in the Kac-Moody invariant theories.
Extremal branes in   more non-compact dimensions differ from the
one dimensional ones only by the dependance
of a harmonic function on the number of non-compact dimensions. For
such  decompactified solutions to exist in
the Kac-Moody theory,   higher level fields must provide the
derivatives needed to obtain  higher dimensional harmonic
functions. This test is crucial to settle the issue of whether or not
the Kac-Moody theories discussed here can really describe uncompactified
space-time covariant theories.

The paper is organised as follows. In section 2, we recall  the
construction of the $\G$-invariant actions $S_{\G}$.  In section 3, we
show how the `cosmological' $\g$-invariant action $S_{{\cal G}_C^{++}}$
is obtained from $S_{\G}$ by a consistent truncation.  In section 4, we
obtain the
`brane' action $S_{{\cal G}_B^{++}}$ which arises from a Weyl
transformation in $\G$ and a consistent truncation.
We show how to formulate $S_{{\cal G}_B^{++}}$ in various equivalent
ways to exhibit various Lorentzian signatures. Differential
equations for the Weyl  transformations of fields ensuring the
equivalence of the different formulations are obtained. The equivalence 
proves that, in addition to 
conventional ones, exotic dualities are present in  $\G$. This is made
explicit in the particular case of $E_{11}$ and is  in agreement
with the results of reference \cite{keu1, keu2}, linking in
$\G$ M-theory to M' and M*-theories. In Section 5, we derive the exact solutions
of $S_{{\cal G}_B^{++}}$, and hence of $S_{\G}$, which are identical to the
space-time solutions describing intersecting extremal branes, conventional and
exotic, smeared in all directions but one. We illustrate by a specific
non-trivial example how dualities transforming these solutions into one
another operate. In Section 6 we stress  perspectives suggested by our
results.

\section{$\G$-invariant action}

Actions $S_{\G}$ invariant under non-linear
transformations of
$\G$ are constructed recursively from a level
decomposition with
respect to a subalgebra
$A_{D-1}$ where $D$ is interpreted as the space-time dimension. The
action is defined in a reparametrisation invariant way  on a
world-line, a priori unrelated to space-time, in terms of fields
$\varphi(\xi)$ where $\xi$ spans the world-line. The fields $\varphi(\xi)$
live in a coset space
$\G/K^{+++}$ where the subalgebra $K^{+++}$ is invariant under a
`temporal involution'
 preserving at each level a Lorentz algebra\footnote{In Section 4, we shall see
that $G^{+++}$  contains other $A_{D-1}$ subalgebras intersecting
differently with
$K^{+++}$.}
$SO(1,D-1)= A_{D-1} \cap K^{+++}$.

We now recall in more detail the construction of these
$\G$-invariant theories
\cite{eh}.

$\G$ contains a  subalgebra $GL(D)$ such that $SL(D) (=A_{D-1})
\subset GL(D) \subset
\G$. The generators of the $GL(D)$ subalgebra are taken to be
$K^a_{~b}\ (a,b=1,2,\ldots ,D)$   with commutation relations
\begin{equation}
\label{Kcom} [K^a_{~b},K^c_{~d}]   =\delta^c_b
K^a_{~d}-\delta^a_dK^c_{~b}\,  .
\end{equation}  The $K^a_{~b}$ along with the  abelian generator $R$,
which is present  when the corresponding maximally oxidised action
$S_{\cal G}$ has one
dilaton\footnote{All the maximally oxidised  theories have at most one dilaton except
the
$C_{q+1}$-series.  The maximally oxidised theory
$C_{q+1}$ is a four dimensional theory   containing $q$ dilatons and
 $C_{q+1}^{+++}$  is constructed with
$q$ abelian generators $R_i \quad i=1 \dots q$ (see for instance ref
\cite{eh} appendix A3). To avoid crowding of indices, we consider
here  only the theories with at most one dilaton.}, are the level zero
generators. The step operators of level greater than zero are tensors
$R^{\quad c_1\dots c_r}_{ d_1\dots d_s}$ of the
$A_{D-1}$ subalgebra. The lowest levels contain antisymmetric tensor
step  operators
$R^{a_1a_2 \dots a_r}$ associated with  electric and magnetic
roots arising from the dimensional reduction of field strength forms in $S_{\cal
G}$. They satisfy the tensor and scaling relations
\begin{eqnarray}
\label{root} &&[K^a_{~b},R^{a_1\dots a_r}]   =\delta^{a_1}_b R^{aa_2
\dots a_r} +\dots +
\delta^{a_r}_b R^{a_1 \dots a_{r-1}a}\, ,\\
\label{root2} && [R, R^{a_1\dots a_r}
] =   -\frac{\varepsilon_A
a_A}{2}\,  R^{a_1\dots a_r}\, ,
\end{eqnarray} where $a_A$ is the dilaton coupling constant to the
 field strength form   and
$\varepsilon_A$ is $+1\, (-1)$ for an
electric (magnetic) root \cite{ehtw}. The generators obey the invariant
scalar product relations
\begin{eqnarray}
\label{killing}
&&\langle K_{~a}^a K_{~b}^b\rangle =G_{ab}\, ,\quad \langle
K^b_{~a}K_{~c}^d\rangle=
\delta_c^b\delta_a^d \ a\neq b\, , \quad\langle R R\rangle
=\frac{1}{2}
\,,\\
\label{step}
&&\langle
R^{\quad a_1\dots a_r}_{ b_1\dots b_s} , \bar  R_{ d_1\dots   d_r}^{\quad
c_1\dots c_s}\rangle
=\delta^{c_1}_{b_1}\dots\delta^{c_s}_{b_s}\delta^{a_1}_{d_1}\dots\delta^
{a_r}_{d_r}\,.
\end{eqnarray}
Here $G= I_D -
\frac{1}{2}\Xi_D$ where $\Xi_D$ is a D-dimensional matrix with all
entries   equal to unity and   $\bar  R_{ d_1\dots   d_r}^{\quad
c_1\dots c_s}$ designates the negative step operator conjugate to $ R^{\quad
d_1\dots   d_r}_{ c_1\dots c_s}$.

The temporal involution $\Omega_1$ generalises the Chevalley
involution to allow identification of the index 1 to a time coordinate in
$SO(1,D-1)$. It is  defined by
\begin{equation}
\label{map} K^a_{~b}\stackrel{\Omega_1}{\mapsto}
-\epsilon_a\epsilon_b K^b_{~a}\quad R\stackrel{\Omega_1}{\mapsto} -R\quad
, \quad R^{\quad c_1\dots c_r}_{ d_1\dots d_s}
\stackrel{\Omega_1}{\mapsto}
-\epsilon_{c_1}\dots\epsilon_{c_r}\epsilon_{d_1}\dots\epsilon_{d_s}
  \bar R_{ c_1\dots c_r}^{\quad d_1\dots d_s}\, ,
\end{equation}
 with $\epsilon_a =-1$ if $a=1$ and
$\epsilon_a=+1$ otherwise. It leaves invariant a subalgebra $K^{+++}$ of
$\G$.

 The fields $\varphi(\xi)$ living in the coset space ${\G}/K^{+++}
$ parametrise the Borel group  built out of Cartan and positive
step operators in
$\G$. Its elements $\cal V$  are written as
\begin{equation}
\label{positive} {\cal V(\xi)}= \exp (\sum_{a\ge b}
h_b^{~a}(\xi)K^b_{~a} -
\phi(\xi) R) \exp (\sum
\frac{1}{r!s!} A^{\quad a_1\dots a_r}_{ b_1\dots b_s}(\xi) R_{
a_1\dots   a_r}^{\quad b_1\dots b_s} +\cdots)\, ,
\end{equation}
 where the first exponential contains only  level zero  operators and
the second one the positive step operators of levels strictly greater
than zero. Defining
\begin{equation}
\label{sym}
dv(\xi)= d{\cal V} {\cal V}^{-1}\quad d\tilde
v(\xi)=   -\Omega_1
\, dv(\xi)
\qquad\quad dv_{sym}=\frac{1}{2} (dv+d\tilde v)\, ,
\end{equation}
one obtains, in terms of the
$\xi$-dependent fields, an  action $S_{\cal \G}$ invariant under
global $\G$ transformations, defined on  the
coset ${\G}/K^{+++}$
\begin{equation}
\label{actionG} S_{\cal \G}=\int d\xi  \frac{1}{n(\xi)}\langle
(\frac{dv_{sym}(\xi)}{d\xi})^2\rangle\, ,
\end{equation} where
$n(\xi)$ is an arbitrary lapse function ensuring reparametrisation
invariance on the world-line. One has
\begin{equation}
\label{sumsym}
dv_{sym}= dv_{sym}^0+\sum_A dv_{sym}^{(A)}\, ,
\end{equation}
where $dv_{sym}^0$ contains all the level zero contributions. One gets
\begin{equation}
\label{v0} dv_{sym}^0=-\frac{1}{2}\sum_{a\ge b} [e^  h (de^{-
h})]_b^{~a} (K^b_{~a}- \Omega_1 K^{b}_{~a}) -
d\phi  R\, ,
\end{equation}
\begin{equation}
\label{vform} dv^{(A)}_{sym}=\frac{1}{2r!s!}DA_{\mu_1\dots
\mu_r}^{\quad
\nu_1\dots
\nu_s} \, \exp (-
\lambda \phi)\, e^{~\mu_1}_{{a}_1}...
e^{~\mu_r}_{{a}_r}e_{\nu_1}^{~{b}_1}
... e_{\nu_s}^{~{b}_s}\, (R^{\quad a_1\dots a_r}_{ b_1\dots
b_s}-\Omega_1 R^{\quad a_1\dots a_r}_{ b_1\dots
b_s})\,  ,
\end{equation}
where $e_\mu^{~a}=(e^{-h(\xi)})_\mu^{~a}$, $\lambda$ is the
generalisation of the scale parameter
$-\varepsilon_A a_A/2$ to all roots
and $D/D\xi$ is a  covariant derivative generalising
$d/d\xi$ through  non-linear terms arising from  non-vanishing
commutators  between  positive  step operators.

Writing
\begin{equation}
\label{full} S_{{\cal G}^{+++}} =S_{{\cal G}^{+++}}^{(0)}+\sum_A
S_{{\cal G}^{+++}}^{(A)}\, ,
\end{equation} where $S_{{\cal G}^{+++}}^{(0)}$ contains all level
zero contributions, one obtains
\begin{equation}
\label{fullzero} S_{{\cal G}^{+++}}^{(0)} =\frac{1}{2}\int d\xi
\frac{1}{n(\xi)}\left[\frac{1}{2}(g^{\mu\nu}g^{\sigma\tau}-
\frac{1}{2}g^{\mu\sigma}g^{\nu\tau})\frac{dg_{\mu\sigma}}{d\xi}
\frac{dg_{\nu\tau}}{d\xi}+
\frac{d\phi}{d\xi}\frac{d\phi}{d\xi}\right],
\end{equation}
\begin{equation}
\label{fulla} S_{{\cal G}^{+++}}^{(A)}=\frac{1}{2 r! s!}\int d\xi
\frac{ e^{- 2\lambda
\phi}}{n(\xi)}\left[
\frac{DA_{\mu_1\dots \mu_r}^{\quad \nu_1\dots
\nu_s}}{d\xi} g^{\mu_1{\mu}^\prime_1}...\,
g^{\mu_r{\mu}^\prime_r}g_{\nu_1{\nu}^\prime_1}...\,
g_{\nu_s{\nu}^\prime_s}
\frac{DA_{{\mu}^\prime_1\dots {\mu}^\prime_r}^{\quad
{\nu}^\prime_1\dots {\nu}^\prime_s}}{d\xi}\right].
\end{equation}  The $\xi$-dependent fields $g_{\mu\nu}$ are defined as
$g_{\mu\nu} =e_\mu^{~a}e_\nu^{~b}\eta_{ab}$. The appearance of the
Lorentz metric $\eta_{ab}$ with $\eta_{11}=-1$ is a consequence of the
temporal involution $\Omega_1$. The metric $g_{\mu\nu}$ allows a
switch from  the Lorentz  indices  $(a,b)$ of the fields appearing  in
Eq.(\ref{positive}) to
$GL(D)$ indices $(\mu,\nu)$.

Exact solutions have been found for all $\G$ theories. In \cite{eh}
solutions describing the algebraic properties of the BPS extremal
branes have been discovered.  Each single extremal p-brane is
characterised by  {\it one\/} non-zero field multiplying a  component
of an antisymmetric tensor step operator of low level.  Each extremal
brane is related in this way to a  {\em real positive root} in $\G$. In
\cite{eh2} these results have been extended and exact solutions of
$S_{\G}$ for all  the extremal intersecting brane solutions
\cite{aeh,ah} of the maximally oxidised theory  were found. The
intersection rules determining such  configurations are neatly encoded
in the $\G$ algebra. They are expressed as
orthogonality  conditions on the real roots characterising  the intersecting
branes.

\setcounter{equation}{0}
\section{From $\G$ to  the cosmological $\g_C$-invariant action }

Consider the overextended algebra $\g_C$ obtained from the very-extended
algebra $\G$  by deleting the  node labelled 1 from the Dynkin
diagrams of $\G$ depicted in Fig.1. This algebra is realised in the space of
Kasner solutions of the action $S_{\G}$ and more generally in the
cosmological billiards solutions\footnote{In what follows, we use the notation
$\g$ when referring to the multiplication table of the overextended algebra and
add an index to $\g$  when we want to keep track of the
embedding in $\G$.}. In this
section we show how  an action invariant under
$\g_C$  emerges from the
$\G$-invariant action $S_{\G}$   Eq.(\ref{full}) through a {\it
consistent} truncation. This `cosmological' action $S_{\g_C}$ generalises to
all $\g$  the
$E_{10}\, (\equiv E_8^{++})$ action of reference \cite{damourhn02}. It is 
obtained below from
$S_{\cal
\G}$ by putting  to zero  in the coset representative Eq.(\ref{positive}) the
field multiplying the Chevalley generator
$H_1=  K_{~1}^1- K_{~2}^2$ and all the
fields multiplying the positive step operators associated to  roots whose
decomposition in terms of simple roots contains  the deleted root
$\alpha_1$.  Such a truncation of the action
$S_{\cal \G}$ will be shown to be consistent with all its equations of motion.

Consider first the fields $A^{\quad a_1\dots a_r}_{ b_1\dots
b_s}(\xi)$ multiplying the positive step
operators $R_{
a_1\dots   a_r}^{\quad b_1\dots b_s}$.  The $A^{\quad a_1\dots a_r}_{
b_1\dots b_s}(\xi)$ equated to zero fall into two classes.
First the set of all  fields appearing in  irreducible
representations of $A_{D-1}$ whose lowest weight contains $\alpha_1$.
Second, in the other representations of
$A_{D-1}$, the set of  tensor components with at least one time index. The
consistency of the truncation of the fields from the first class is obvious as
all  terms in the action Eq.(\ref{full}) either contain no such field or
contain at least two of them. Fields of the second class may occur linearly but
are then multiplied by non diagonal metrics
$g_{1\hat{\mu}}$ where the hatted indices  $\hat{\mu}$ run from
$2$ to $D$. In the triangular gauge for $h_\mu^{~a}$ stemming from the Borel
group defined in Eq.(\ref{positive}), the linear terms in the second class
$A$-fields are necessarily multiplied by at least
one $h_1^{~a} $-field $(a>1)$ associated through $K^1_{~a}$ to
a root $\alpha_1$. Taking
\begin{equation}
\label{ofdiag} g_{1\hat{\mu}}=0\, ,
\end{equation}
we ensure the consistency of putting to zero the fields of the second
class and in fact of all the fields multiplying step operators associated
to a root
$\alpha_1$, including those of level zero. Indeed, the equation of
motions of the second class $A$-fields are now satisfied. Consider now
the equations of motion of the $g_{1\hat{\mu}}$. All the terms
containing $g_{1\hat{\mu}}$  contain necessarily at least
one other
$g_{1\hat{\nu}}$ in Eq.(\ref{fullzero}) or at least one 
$A$-field component put to zero in
Eq.(\ref{fulla}). The equations of motion of the $g_{1\hat{\mu}}$
are thus also consistent with the truncation.

We now turn to the  fields multiplying Cartan generators. Labelling  $q_1$
 the field multiplying $H_1$ in the
Chevalley basis we express the equation $q_1=0$ in terms of the metric
$g_{\mu\nu}$. We define
\begin{equation}
\label{pdef}
p_a =  - h^{~a}_a\, ,
\end{equation}
where $h^{~a}_a$ is the
coefficient of $K^a_{~a}$ in Eq.(\ref{positive}) . One gets for all $\G$
\cite{ehtw} the equivalence
\begin{equation}
\label{embed}
 q_1=0\qquad\Longleftrightarrow\qquad
 p_1 = \sum_{a=2}^{D} p_a\, .
\end{equation}  Using the fact that the coset model Eq.(\ref{actionG})
was computed in  the triangular Borel  gauge and using Eq.(\ref{ofdiag}),
we  rewrite Eq.(\ref{embed}) as
\begin{eqnarray}
\label{det1} g_{11}&=&{\bf g}\, ,\\
\label{det2} {\bf g}&\equiv&\det{g_{\hat{\mu}\hat{\nu}}}\, .
\end{eqnarray}

Consistency of the truncation from $\G$ to $\g$ requires
that Eq.(\ref{det1}) satisfy the equation of motion of
$g_{11}$. Only variations with respect to $g_{11}$ in the action
Eq.(\ref{fullzero}) have to be considered, as the
variations in Eq.(\ref{fulla}) are automatically satisfied because
$g_{11}$ multiplies  $A$-fields already equated to zero.
Using Eq.(\ref{ofdiag})  we get
\begin{equation}
\label{eomzero}
\frac{d^2}{d\xi^2}\ln |g_{11}|-\frac{d^2}{d\xi^2}\ln{\bf g} = 0\, ,
\end{equation} which admits  as solution Eq.(\ref{det1}).

To obtain the  actions $S_{\g_C}$  whose
equations of motion are all contained in the equations of motion
obtained from
$S_{\G}$ given by Eq.(\ref{actionG}), it thus suffices to substitute in
Eq.(\ref{fullzero}) $g_{11}$ by its value given in Eq.(\ref{det1}) and
equate to zero in Eq.(\ref{fulla}) all first and second
class $A$-fields. Denoting the remaining $A$-fields by $B$ and using the
hatted indices $\hat\mu , (\mu=2,\dots D)$, we obtain

\begin{equation}
\label{fullp} S_{\g_C} = S_{\g_C}^{(0)}+\sum_B
S_{\g_C}^{(B)}\, ,
\end{equation} where
\begin{eqnarray}
\label{fullop} S_{\g_C}^{(0)}& =&\frac{1}{2}\int dt
\frac{1}{n(t)}\left[\frac{1}{2}(g^{\hat{\mu}\hat{\nu}}g^{\hat{\sigma}\hat{\tau}}-
g^{\hat{\mu}\hat{\sigma}}g^{\hat{\nu}\hat{\tau}})\frac{dg_{\hat{\mu}
\hat{\sigma}}}{d\tau}
\frac{dg_{\hat{\nu}\hat{t}}}{dt}+
\frac{d\phi}{dt}\frac{d\phi}{dt}\right] ,\\
\label{fullc} S_{\g_C}^{(B)}&=&\frac{1}{2 r! s!}\int dt
\frac{ e^{- 2\lambda
\phi}}{n(t)}\left[
\frac{DB_{\hat{\mu}_1\dots \hat{\mu}_r}^{\quad \hat{\nu}_1 \dots
\hat{\nu}_s}}{dt} g^{\hat{\mu}_1{\hat{\mu}}^\prime_1}...\,
g^{\hat{\mu}_r{\hat{\mu}}^\prime_r}g_{\hat{\nu}_1{\hat{\nu}}^\prime_1}
...\ ,    g_{\hat{\nu}_s{\hat{\nu}}^\prime_s}
\frac{DB_{{\hat{\mu}}^\prime_1\dots {\hat{\mu}}^\prime_r}^{\quad
{\hat{\nu}}^\prime_1\dots {\hat{\nu}}^\prime_s}}{dt}\right] .
\end{eqnarray} Here we  renamed $\xi$ as $t$.
Eqs.(\ref{fullp}),(\ref{fullop}) and(\ref{fullc}) follow from 
Eq.(\ref{actionG}) by replacing
$dv_{sym}$ by
$dv{_{sym}}_{(+)}$,   which describes a motion on the coset
$\g/K^{++}_{(+)}$ where  $K^{++}_{(+)}$ is the subalgebra of $\g$
invariant under the Chevalley involution. In particular, we
see that for $\g = E_8^{++}$, we recover as a consistent truncation of
$S_{E_8^{+++}}$  given by Eq.(\ref{actionG}) the action for the
$E_{10}\equiv E_8^{++}$-invariant theory proposed in reference
\cite{damourhn02}.

The fact that the solutions of the $\g$-equations of motion are
also solutions of the $\G$-equations of motion corresponding to a
particular choice of the initial conditions can be viewed in a
different way. Consider the general solution of the $\G/K^{+++}$
non linear sigma-model. In the notations of \cite{damourhn00}, it
reads
\begin{equation} \cm (\xi) = \cm (0) \cdot \exp (\xi J)\, , \label{solu}
\end{equation}
where $\cm$ is defined as $\cm = {\cal V}^{T'} {\cal V}$ in terms
of the $\G$-group element ${\cal V}$.  Here, the generalised
transposition is defined in terms of the involution $\Omega_1$ as
$E^{T'} = - \Omega_1(E)$ for any Lie algebra element $E$ (and
extended by the rule $(AB)^{T'} = B^{T'} A^{T'}$). In
(\ref{solu}), the $J$'s are the conserved currents and belong
to $\G$.  If we choose the initial conditions in such a way
that $\cm (0)$ belongs to the $\g$-subgroup and the current $J$
is in the subalgebra $\g$, then we see from (\ref{solu}) that
$\cm (\xi)$ is in the $\g$-subgroup for all ``times" $\xi$.
This proves consistency of the truncation at the level of the
equations of motion.

\setcounter{equation}{0}
\section{From $\G$ to  the brane $\g_B$-invariant action}

\subsection{Weyl reflections of the gravity line}
The Weyl reflection $W_{\alpha_1}$ in the hyperplane perpendicular to
$\alpha_1$ generates a subalgebra $\g_2$ conjugate to $\g_C$ in
$\G$. More generally, performing  Weyl reflections from roots of the
gravity line, we generate in this way $(D-1)$  subalgebras $\g_a,
(a = 2,\dots, D)$, which are all conjugate in $\g$.  While the generators of
$\g_C$ were labelled by  spatial indices only, the
index
$a$ in
$\g_a$ must   be interpreted as  a time coordinate. The fact
that we have $D-1$ different identifications of the  time
coordinate follows from the non-commutativity of  the time
involution with the Weyl reflections \cite{keu1} as seen
below. This non-commutativity will be exploited in the 
more general context of Section 4.2 but is well illustrated by
its effect on the gravity line.

Expressing a Weyl transformation $W$  as a conjugation by a group
element
$U_W$ of $\G$, we define the involution $\Omega^\prime$ operating on
the conjugate elements by
\begin{equation}
\label{newinvolve}
U_W\, \Omega T\, U^{-1}_W=\Omega^\prime \, U_W  T U^{-1}_W\, ,
\end{equation}
where $T$ is any generator of $\G$.  Applying Eq.(\ref{newinvolve}) to
the Weyl reflection $W_{\alpha_1}$  generates the subalgebra
$\g_2$ conjugate to
$\g_C$. One gets
\begin{eqnarray}
\label{permute}
&&U_1\, \Omega K^2_{\ 1} \, U^{-1}_1= \rho K^2_{\ 1}= \rho\Omega^\prime
 K^1_{\ 2}\nonumber\, ,\\
&&U_1\, \Omega K^1_{\ 3} \, U^{-1}_1= \sigma K^3_{\ 2}=
\sigma\Omega^\prime
 K^2_{\ 3} \, ,\\
&&U_1\, \Omega K^i_{\ i +1} \, U^{-1}_1= -\tau K^{i+1}_{\ \, i}=
\tau\Omega^\prime  K^i_{\ i +1}\quad i >2\, .\nonumber
\end{eqnarray}
Here $\rho,\sigma,\tau$ are plus or minus signs which may arise
as step operators are representations of the Weyl group 
up to signs. Eq.(\ref{permute}) illustrate the general result
that such signs always cancel in the determination of
$\Omega^\prime$ because they are identical in the Weyl transform of
corresponding positive and negative roots, as
their commutator is in the Cartan subalgebra which  forms a true
representation of the Weyl group. The content of Eq.(\ref{permute}) is
represented in Table 1. The signs below the generators of the gravity
line indicate the sign in front of the
 negative step operator obtained by the involution: a
minus sign is in agreement with the conventional Chevalley involution
and indicates that the indices in
$K^a_{\ a +1}$ are both either space or time indices while a plus sign
indicates that one index must be time and the other  space.
\begin{table}[h]
\caption{\small Involution switch from $\Omega$ to
$\Omega^\prime$ due to the Weyl reflection $W_{\alpha_1}$}
\begin{center}
\begin{tabular}{|c|ccccc|c|}
\hline
gravity line&$K^1_{\ 2}$&$K^2_{\ 3}$&$K^3_{\ 4}$&$\cdots$&$K^{D-1}_{\
D}$&time coordinate\\
\hline\hline
$\Omega$&$+$&$-$&$-$&$-$&$-$&1\\
\hline$\,\Omega^\prime$&$+$&$+$&$-$&$-$&$-$&2\\
\hline
\end{tabular}
\end{center}
\label{gravity}
\end{table}

\noindent
Table 1 show that
the  time coordinates in
$\g_2$ must now be identified either with 2, or with all indices
$\neq 2$. We choose the first description, which leaves
unaffected coordinates attached to planes invariant under the Weyl
transformation. Similarly the time coordinate in
$\g_a$ is taken to be
$a$.

One may now apply the analysis of Section 3 to obtain  actions
$S_{\g_a}$ invariant under $\g_a$, namely we equate as before to zero
all the fields in the $\G$-invariant action Eq.(\ref{full}) which
multiply generators not involving the root $\alpha_1$, but this
truncation is performed {\em after} the Weyl transformation which
transmutes the time index 1 to a space index. The actions $S_{\g_a}$ are then
formally identical to the one given by Eqs.(\ref{fullp}),
(\ref{fullop}) and (\ref{fullc}) but with a Lorentz signature for the
metric, which in the flat coordinates amounts to a negative sign for the
Lorentz metric component $\eta_{aa}$, and with $\xi$ identified to the missing
space coordinate instead of $t$.

The $D-1$ actions $S_{\g_a}$ for $a = 2,\dots, D$  differ only by the
index identifying the time coordinate and the concomitant identification of
$\xi$ with a space coordinate. They are thus equal up to a trivial redefinition
of all tensor fields by an interchange of indices. This redefinition may be
viewed as the Weyl transformations generated by the roots of the subalgebra
$A_{D-2}$ on the space of fields. Equivalence under Weyl transformations from
roots which do not belong to the gravity line are far less trivial and will now
be examined.

\subsection{General Weyl reflections}

The $D-1$  actions $S_{\g_a}$ differ by the labelling of the
time coordinate and have all the same global signature $(1,9)$.   They
are related through Weyl transformations of $\g$ from roots of
the gravity line. Consider now the Weyl transformations of $\g$
generated in addition by the simple roots not belonging to $A_{D-2}$.
These will yield  actions with different global signatures.
All such actions will be shown to be equal and related by field
redefinitions. This equivalence realises in the action
formalism  the general analysis of Weyl transformations by Keurentjes
\cite{keu1,keu2}. It will be related to the existence, in
addition to conventional duality symmetries, to exotic dualities, which
in the particular case of M-theory, are relating it to M* and
M'-theories \cite{hull1,hull2}. To illustrate how Weyl transformation change
global signatures, we first consider explicitly  the signature changes for
$E_8^{++}\equiv E_{10}$.

The Weyl reflection from the root
associated to any generator $R^{abc}$ multiplying the three-form
potential $A_{abc}$ can always be mapped to the Weyl reflection
$W_{\alpha_{11}}$ from the simple root $\alpha_{11}$ by products of
Weyl reflections of the gravity line, thereby changing the  time
coordinate, originally positioned at 1, to any position.

Let us choose the time coordinate to be 9 and consider the Weyl
reflection $W_{\alpha_{11}}$. The only generator of the gravity line
affected by the Weyl transformation is $K^8_{\ 9}$. One has from
Eq.(\ref{newinvolve})
\begin{equation}
\label{weyla}
U_{11}\, \Omega R^{8\,10\,11} \, U^{-1}_{11}= -\rho K^9_{\ 8}=
\rho\Omega^\prime
 K^8_{\ 9}\, .
\end{equation}
Imposing that the coordinate 2 remains unaffected by the transformation,
we see that both 10 and 11 become time coordinates, as
illustrated in Table 2. The transformation of the  generator
$R^{9\,10\,11}$ yields
\begin{equation}
\label{weylb}
U_{11}\, \Omega R_{9\,10\,11} \, U^{-1}_{11}= \sigma R_{9\,10\,11}=
\sigma\Omega^\prime
 R^{9\,10\,11}.
\end{equation}
The action of the involution
$\Omega^\prime$  on the simple root  not
pertaining to the gravity line, $\Omega^\prime R^{9\,10\,11}=
R_{9\,10\,11}$, differs by a sign from the action of a temporal
involution  defined according to Eq.(\ref{map}) when the times are
10 and
11.   This shift of sign is shown
in the last column of Table~\ref{e10}. It will lead to negative kinetic
energy terms in the corresponding actions
$S_{{\cal G}{}_{(10\,11)}^{++}}^{(2,8,-)}$ below. Table~\ref{e10}
shows how by repeated use of Eqs.(\ref{weyla}) and (\ref{weylb}) one
reaches the signatures $(5,5,+)$,
$(6,4,-)$ and
$(9,1,+)$.

\begin{table}[h]

\caption{\small Involution switches from $\Omega$ to
$\Omega^\prime$ due to the Weyl reflection $W_{\alpha_{11}}$}

\begin{center}
\begin{tabular}{|ccccccccc|c|c|}
\hline
$K^2_{\ 3}$&$K^3_{\ 4}$&$K^4_{\ 5}$&$K^5_{\ 6}$&$K^6_{\
7}$&$K^7_{\ 8}$&$K^8_{\ 9}$&$K^9_{\ 10}$&$K^{10}_{\ 11}$&
times & $(t,s,\pm)$ \\
\hline\hline
$-$&$-$&$-$&$-$&$-$&$-$&$+$&$+$&$-$&9&$(1,9,+) $\\
\hline $-$&$-$&$-$&$-$&$-$&$-$&$-$&$+$&$-$&10,11&$(2,8,-)
$\\
\hline\hline
$-$&$-$&$-$&$-$&$+$&$-$&$+$&$-$&$-$&7,8&$(2,8,-) $\\
\hline $-$&$-$&$-$&$-$&$+$&$-$&$-$&$-$&$-$&7,8,9,10,11&$(5,5,+)
$\\
\hline\hline$-$&$-$&$-$&$+$&$-$&$-$&$+$&$+$&$-$&2,3,4,5,9&$(5,5,+)
$\\
\hline$-$&$-$&$-$&$+$&$-$&$-$&$-$&$+$&$-$&2,3,4,5,10,11&$(6,4,-)
$\\
\hline\hline$-$&$-$&$-$&$-$&$-$&$+$&$-$&$-$&$-$&2,3,4,5,6,7&$(6,4,-)
$\\
\hline$-$&$-$&$-$&$-$&$-$&$+$&$+$&$-$&$-$&2,3,4,5,6,7,9,10,11&$(9,1,+)
$\\
\hline
\end{tabular}
\end{center}
\label{e10}
\end{table}

It is interesting to illustrate the string interpretation of the
Weyl transformations of Table 2  in terms of signature changing
dualities discussed in \cite{hull1,hull2}. Consider the type IIA interpretation
of $E_8^{+++}$. In the Dynkin diagram of
$E_8^{+++}$ in Fig. 1 the node labelled 10 is no longer on
the gravity line and should be redrawn as perpendicular to it at the node
labelled 9. Recall that in the embedding of  $E_8^{++}$ in
$E_8^{+++}$ the coordinate 1 is  a space coordinate and that the string
interpretation of the Weyl reflection $W_{\alpha_{11}}$ is a double
$T$-duality in the directions 9 and 10 plus an exchange
of these  directions \cite{weylt,banks,ehtw}.

The first Weyl
reflection of Table 2  maps  the
$E_8^{++}$ signature
$(1,9,+)$ to $(2,8,-)$. To get the corresponding string theories we  ignore the
direction $11$ and take into account the direction
$1$ which is spacelike. The $(1,9,+)$ signature corresponds to type
$IIA$ theory. 
The action of $W_{\alpha_{11}}$ renders 11  timelike. Consequently the
string interpretation of the $(2,8,-)$ signature is a string theory with
signature $(1,9)$ and wrong signs for the kinetic terms in the $RR$ sector.
This defines the type $IIA^*$ theory consistently  with 
string dualities. Indeed a double T-duality with respect to one time
direction, 9, and one space direction, 10, maps type $IIA$ onto
type $IIA^*$ \cite{hull1,hull2}. The second Weyl
reflection of Table 2  maps  the
$E_8^{++}$ signature
$(2,8,-)$ with  time directions $7,8$ to $(5,5,+)$ with time directions
$7,8,9,10,11$. The string interpretation is obtained by the same argument
as above and one gets in this way a mapping of  $IIA_{8+2}$ to
$IIA_{6+4}$ theories, in the notation of
\cite{hull2}, where the index $s+t$ designates a theory with $s$ space
 and $t$ time directions. This result is in agreement with the string
dualities. Indeed a double spacelike
$T$-duality in the direction $9,10$ maps $IIA_{8+2}$ to
$IIA_{6+4}$ (see Figure 5 of \cite{hull2}). The third Weyl
reflection of Table 2  maps  the
$E_8^{++}$ signature
$(5,5,+)$ with  time directions $2,3,4,5,9$ on $(6,4,-)$ with time
directions
$2,3,4,5,10,11$. The string interpretation maps accordingly type $IIA_{5+5}$
to  type $IIA^*_{5+5}$. These two theories are indeed related by a double
$T$-duality in the time direction 9 and in the space direction
10. The last Weyl reflection of Table 2 maps the $E_8^{++}$ signature
$(6,4,-)$ with  time directions $2,3,4,5,6,7$ to $(9,1,+)$ with time
directions
$2,3,4,5,6,7,9,10,11$, in accordance with     the string interpretation
where type $IIA_{4+6}$ is mapped to type $IIA_{2+8}$ by
double spacelike T-duality.

\subsection{Weyl equivalence of different space-time
signatures}

The action $S_{\g_2}$ is, as $S_{\g_C}$, given by
Eqs.(\ref{fullp}),(\ref{fullop}) and (\ref{fullc}) but with
$\eta_{22}=-1$. It can be expressed in the general form 
Eq.(\ref{actionG}) by replacing
$dv_{sym}$ by $dv{_{sym}}_{(-)}$   which describes a motion on the coset
$\g/K^{++}_{(-)}$.  $K^{++}_{(-)}$ is the subalgebra of $\g$
invariant under the  time involution $\Omega_2$ defined as in
Eq.(\ref{map}) with 2 as the time coordinate and restricted to $\g$. To
compute
$dv{_{sym}}_{(-)}$ one specifies a
coset  representative of $\g/K^{++}_{(-)}$ and a level expansion
about a gravity line endowed with the involution
$\Omega_2$. We shall refer to this embedding of $\g$ in $\G$ by $\g_B$
to distinguish it from $\g_C$.

The action $S_{\g_2}$  is characterised by a global signature $(1,9, +)$ and
time coordinate~2. We may reparametrise the coset space $\g/K^{++}_{(-)}$ by
subjecting its elements to  a Weyl conjugation by an element $U$ of $\g$.
This selects a new gravity line in $\g$ which may be endowed with a
different involution than $\Omega_2$, as the temporal involution does
not in general commute with Weyl transformations. The Borel group and the
level expansion accommodating the new involution will then not coincide with
the old one, although the new $\xi$-dependant fields could
always in principle be expressed in terms of the old ones. One can
construct in this way apparently many distinct actions in terms of these
different set of fields.  Weyl
transformations from the gravity line roots permute the tensor indices and
hence may change  time indices while Weyl reflections from roots which do
not belong to the gravity line may change  the global signature, as
exemplified for
$E_8^{++}$ in Table 2.  We denote these actions by
$S_{{\cal G}{}_{(i{}_1i{}_2\dots i{}_t)}^{++}}^{(t,s,\varepsilon)}$, where
the global signature is $(t,s,\varepsilon)$ with
$\varepsilon$ denoting  a set of $+1$ or $-1$ signs associated to each
simple root which does not pertain to the gravity line, and
$i{}_1i{}_2\dots i{}_t$ are the time indices. We shall verify that all these
actions are equivalent and that their corresponding Lagrangians are equal
for all $\xi$. We show the equivalence by deriving the differential equations
 relating the fields parametrising the different coset representatives.

We write  $S_{{\cal G}{}_{(i{}_1i{}_2\dots
i{}_t)}^{++}}^{(t,s,\varepsilon)}$  in terms of
$dv_{sym \, (i{}_1i{}_2\dots i{}_t)}^{(t,s,\varepsilon)}$ as
\begin{equation}
\label{newaction}
 S_{{\cal G}{}_{(i{}_1i{}_2\dots
i{}_t)}^{++}}^{(t,s,\varepsilon)} =\int d\xi  \frac{1}{n(\xi)}\langle
(\frac{dv_{sym \, (i{}_1i{}_2\dots
i{}_t)}^{(t,s,\varepsilon)}(\xi)}{d\xi})^2\rangle\, ,
\end{equation}
where $dv_{sym \, (i{}_1i{}_2\dots i{}_t)}^{(t,s,\varepsilon)}$ is
computed  as in Eqs.(\ref{positive}) and (\ref{sym}), but using a
Borel  representative  of
$\g/K^{++}_{(-)}$,  built  from a gravity line endowed
with the involution
$\widetilde
\Omega$ defining the signature $(t,s,\varepsilon)$ with
$\{i{}_1i{}_2\dots i{}_t\}$ as time coordinates.
This yields the same expressions as in Eqs.(\ref{fullp}),
(\ref{fullop}) and (\ref{fullc}) but with a different space-time
signature and a change of sign for all terms in levels generated by an
odd number of simple roots not pertaining to the gravity line and
carrying a negative contribution to $\varepsilon$.

We denote the Cartan
generators  $K^a_{~a}$ and $R$  as $H^{
 i}$ and the tensor positive step operators as $R^{
j}$. In terms of the differential forms $\{X_i, Y_j\}$ built from
the Borel group fields , one has
\begin{equation}
\label {new}
 dv_{sym \, (i{}_1i{}_2\dots i{}_t)}^{(t,s,\varepsilon)}=
\sum_i X_i H^i + \sum_j  Y_j
 (R^j -
\widetilde\Omega R^j)\, .
\end{equation}
On the other hand $S_{{\cal G}{}_{(i{}_1i{}_2\dots
i{}_t)}^{++}}^{(t,s,\varepsilon)}$ can be obtained by
conjugation from $S_{\g_2}$ in the following way. Write
$dv{_{sym}}_{(-)}$ as
\begin{equation}
\label{sym-}
dv{_{sym}}_{(-)} = \sum_i X_i^\prime H^{\prime\, i} + \sum_j Y_j^\prime
(R^{\prime\, j} -
\Omega_2 R^{\prime\, j})\, ,
\end{equation}
where $\{X_i^\prime, Y_j^\prime\}$ are the differential forms built   
   from the Borel fields defined by a gravity line endowed with
the involution
$\Omega_2$. Select
 generators $R^{\prime\prime\, j}$  whose  Weyl transform
$R^i=\widetilde U R^{\prime\prime\, j} \widetilde U^{-1},\,
\widetilde U
\subset\g$,
 spans the
positive step generators in the level expansion defining $dv_{sym \,
(i{}_1i{}_2\dots i{}_t)}^{(t,s,\varepsilon)}$ and  maps the
involution
$\Omega_2$ to
$\widetilde \Omega$. We have
\begin{equation}
\label{correspond}
\widetilde U (R^{\prime\prime\, j} -
\Omega_2 R^{\prime\prime\, j}) \widetilde U^{-1}=\rho^j (R^j -
\widetilde\Omega R^j)\, ,
\end{equation}
and $\rho^j$ is a
sign. Note that
$R^{\prime\prime\, j} $ need not be a positive step operator
$R^{\prime\, j}$ in Eq.(\ref{sym-})  but one  always has
\begin{equation}
\label{negative}
 (R^{\prime\prime\, j}  -
\Omega_2 R^{\prime\prime\, j} ) =\lambda^j ( R^{\prime\, j} -
\Omega_2  R^{\prime\, j})\, ,
\end{equation}
  where $\lambda^j$ is a
sign. Performing the conjugation  on $dv{_{sym}}_{(-)}$ one gets
\begin{equation}
\label {conjugate}
\widetilde U dv{_{sym}}_{(-)}\widetilde
U^{-1}= \sum_i X_i^\prime H^i + \sum_j  Y_j^\prime
\lambda^j\rho^j (R^j -
\widetilde\Omega R^j)\, .
\end{equation}
Comparing Eq.(\ref{new}) with Eq.(\ref{conjugate}) we see that the
integrable differential equations relating the different Borel fields
\begin{equation}
\label{field}
\hbox{\framebox{$ X_i^\prime= X_i$\, ,\, $  Y_j^\prime \lambda^j\rho^j=
Y_j$}}\,
\end{equation}
 ensures that $dv_{sym \, (i{}_1i{}_2\dots i{}_t)}^{(t,s,\varepsilon)}$
and $dv{_{sym}}_{(-)}$ define the same Lagrangian for all $\xi$ and
prove the equivalence of
 $S_{{\cal G}{}_{(i{}_1i{}_2\dots
i{}_t)}^{++}}^{(t,s,\varepsilon)}$ and  $S_{\g_2}$.
The resulting field transformations realise
the Weyl transformation $\widetilde U$ in field space. We denote the
set of equivalent actions  by $S_{\g_B}$.
The two distinct  actions $S_{\g_B}$ and $S_{\g_C}$ are both contained
as consistent distinct truncations of the unique action $S_{\G}$, which
encompasses different signatures and can be formulated, as was
done for $S_{\g_B}$, as separate actions identified through field
redefinitions.

We illustrate the field transformations obtained from Eq.(\ref{field}) by
considering for $\g =E_8^{++}$ the equivalence of $S_{{\cal
G}{}_{(9)}^{++}}^{(1,9,+)}$ and $ S_{{\cal G}{}_{(10\,
11)}^{++}}^{(2,8,-)} ,$ under the Weyl transformation $W_{\alpha_{11}}$
given in the first block of Table 2. First consider the non-trivial
relation $ Y_j^\prime \lambda^j\rho^j= Y_j$ at the lowest level. Taking into
account Eq.(\ref{weyla}) and Eqs.(\ref{v0})-(\ref{vform}) we get the following
non-linear relations (up to the sign $\lambda^j\rho^j$ not taken into account here)
\begin{eqnarray}
\label{relaa}
\frac{1}{3!}dA^\prime_{\mu \nu
\rho}\,
 (e^{h^\prime})^{~\mu}_{8} (e^{h^\prime})^{~\nu}_{10}
(e^{h^\prime})^{~\rho}_{11}   &=& -[e^  h
(de^{-h})]_8^{~9}\, ,\\
\label{relab}-[e^{h^\prime}
(de^{-{h^\prime}})]_8^{~9}
 &=&\frac{1}{3!}dA_{\mu \nu
\rho}\, (e^h)^{~\mu}_{8} (e^h)^{~\nu}_{10}
(e^h)^{~\rho}_{11}\, .
\end{eqnarray}
We now turn to the relation $X_j^\prime= X_j$ between the Cartan
generators. From $\sum_{i=2}^{11} X_i^\prime H^{\prime\, i}
\equiv -\sum_{a=2}^{11} dp^{\prime a} K^a_{~a}~(p^{\prime a}=- h^{~\prime a}_a)$ 
one expresses the Weyl reflection
$W_{\alpha_{11}}$ as
\cite{ehtw}
\begin{eqnarray}\nonumber K^{\prime a}{}_a&=&K^a_{~a}\quad a=2,\ldots ,8
\\ \nonumber K^{\prime a}{}_{a}&=&K^a_{~a}+{1\over
3}(K^2_{~2}+\dots+K^8_{~8})-{2\over
3}(K^{9}_{~9}+K^{10}_{~10}+K^{11}_{~11})~~ a=9,10,11 \,
\end{eqnarray} to recover the known result
\begin{eqnarray}
\label{WeylMe2} p^{\prime a}+\frac{1}{3}(p^{\prime
9}+p^{\prime 10}+p^{\prime 11})&=&p^a
\qquad
a=2,\dots, 8\\
\label{WeylMe1} p^{\prime a} -\frac{2}{3}(p^{\prime
9}+p^{\prime 10}+p^{\prime 11}) &=&p^a
\qquad
a=9,10,11\, .
\end{eqnarray}
We shall apply in the next section Eqs. (\ref{relaa}),(\ref{relab}),
(\ref{WeylMe2}), and (\ref{WeylMe1}) to exact 
extremal brane solutions of $\g_B$.

\setcounter{equation}{0}
\section{Intersecting extremal branes in $\g$ theories}
Exact solutions of  $S_{{\cal G}^{+++}}$  giving the algebraic properties
of  intersecting  brane configurations have been constructed in
\cite{eh2}. We  show in this section that in $S_{\g_B}$,
these coincide with the solutions of the Einstein and field equations of
the maximally oxidised theories describing  the intersecting branes
 smeared in all directions but one. $\xi$ is  identified to the
non-compact coordinate.

We consider the level decomposition in $\g_B$ for a  signature
$(t,s,\varepsilon)$ with $D-1=t+s$ connected by  Weyl
reflections  to the `phase' $(1,D-2,+)$ with time coordinate 2 of the
$\g_B$ theory. This is the phase described by the fields in the action
$S_{\g_2}\equiv S_{{\cal G}{}_{(2)}^{++}}^{(1,9,+)}$.

The equations of motion yielding the intersecting brane
configuration solutions from $S_{\G}$ in \cite{eh2} can immediately be
read in $S_{\g_2}$ provided the embedding relation
Eq.(\ref{embed}) is satisfied with  $p_1 = \sum_{a=2}^{D} p_a$ 
labelling  a space direction orthogonal to the branes. The solutions are
unaltered for
$a=2,3\dots D$. As pointed out in
$\cite{eh2}$, these solutions can be extended to exotic branes with
$t_A$ longitudinal timelike and $s_A$ spacelike directions connected in
$\G$ by Weyl reflections. The results of Section 4 yield therefore the
intersecting brane configuration solutions, exotic or not, from the fields
defining 
$S_{{\cal G}{}_{(i{}_1i{}_2\dots
i{}_t)}^{++}}^{(t,s,\varepsilon)}$. For each of the $\cal N$ branes
present in the configuration and characterised by
$\tau_1 \dots \tau_{t_A}$ longitudinal timelike directions and
$\lambda_1\dots
\lambda_{s_A}$ longitudinal spacelike directions, one has
\begin{equation} A_{\tau_1 \dots \tau_{t_A}\lambda_1\dots \lambda_{s_A}}=
\epsilon_{\tau_1 \dots \tau_{t_A}\lambda_1\dots \lambda_{s_A}}
[\frac{2(D-2)}{\Delta_A}]^{1/2}H_A^{-1}(\xi)
\qquad A=1 \dots {\cal N}\, ,
\label{aequ}
\end{equation} and
\begin{eqnarray}
 &&p^a= \sum_{A=1}^{\cal N} p_A^a=\sum_{A=1}^{\cal N}
\frac{\eta_A^a}{\Delta_A}
\ln H_A(\xi)\qquad a=2,3,\dots,D  \label{pequ}\\
\label{phiequ} &&\phi =\sum_{A=1}^{\cal N} \, \phi_A = \sum_{A=1}^{\cal
N}\frac{D-2}{\Delta_A}\varepsilon_A a_A
\ln H_A(\xi) \, .
\end{eqnarray} Here $\eta_A^a=s_A+t_A$ or $-(D-2-s_A-t_A)$ depending on
whether  the direction $a$ is perpendicular or parallel to the
$q_A$-brane and
$\Delta_A= (s_A+t_A)(D-2-s_A-t_A)+\frac{1}{2}a_A^2(D-2)$.  The factor
$\varepsilon_A$ is $+1$ for an electric brane and $-1$ for  a magnetic one. Each
of the  branes in the configuration is thus described as electrically charged and
is characterised by one positive harmonic function  in
$\xi$-space, namely one has
\begin{equation}
\frac{d^2H_A(\xi)}{d\xi^2}=0 \qquad A=1 \dots {\cal N}\, .
\label{harmo}
\end{equation}
From Eq.(\ref{pequ}), the embedding relation Eq.(\ref{embed}) yields for
the spatial  direction 1 the result
\begin{equation}
\label{dir1}
p^1= \sum_{A=1}^{\cal N}
\frac{s_A+t_A}{\Delta_A}\,
\ln H_A(\xi)\, ,
\end{equation}
identifying it to a direction transverse to all branes.
The Eqs. (\ref{pequ}), (\ref{phiequ}) and (\ref{dir1})
are solutions provided the generalised intersection rules
\cite{ah}
\begin{equation}
\bar{s}+\bar{t}=\frac{(s_A+t_A)(s_B+t_B)}{D-2}-\frac{1}{2}
\varepsilon_A a_A \varepsilon_B a_B \label{exorule}
\end{equation}
are satisfied. The intersection rules Eq.(\ref{exorule}) can be expressed
 as an orthogonality condition between the real positive roots of
$\g_B$ (and
$\G$) for all  branes present in the configuration
\cite{eh2}. This orthogonality condition is in fact the input that
permits the derivation of the exact solutions  Eqs.
(\ref{aequ}), (\ref{pequ}),  (\ref{phiequ}) and (\ref{harmo}) by
allowing a reduction of $S_{\g_B}$ to  quadratic
terms.

We may   verify from the above equations  that the  lapse constraint in
$S_{\g_B}$ is satisfied and takes the form
\begin{equation}
\label{xiextremal}
\sum_{a=2}^D ( d p^{a})^2 -(\sum_{a=2}^D
dp^{a})^2 +
\frac{1}{2} (d\phi)^2-  \sum_{A=1}^{\cal N}\frac{D-2}{\Delta_A} (d\ln
H_A)^2=0\, ,
\end{equation}
where the differentials are taken in $\xi$-space.
Eq.(\ref{xiextremal}) expresses the vanishing of the action for
solutions involving fields associated to orthogonal roots. This
condition is preserved under a Weyl reflection, whether or not the
latter induces a signature change. It thus follows from
the invariance of $S_{\g_B}$, as expressed by Eq.(\ref{field}), that the lapse
equation is invariant under  Weyl reflections. In addition the term $\Lambda= -
\sum_{A=1}^{\cal N}(D-2/\Delta_A) (d\ln H_A)^2$ is separately invariant because
the other terms in Eq.(\ref{xiextremal}) are the  quadratic form of $\g$
restricted to its Cartan subalgebra, which is Weyl invariant. This
invariance relates different intersecting branes, Kaluza-Klein momenta
and monopoles \cite{eh,eh2} of different phases, conventional or exotic.

These exact solutions of $S_{\g_B}$ describe  not only the
algebraic properties of all corresponding solutions, exotic or not,
of all the maximally oxidised theories and of their exotic counterparts
related to them by `dualities' encoded in the Weyl group of $\g_B$.
They are now  identical to the solutions of the Einstein and field
equations describing  these
intersecting branes smeared in all directions but the spatial
dimension 1. This follows on the one hand from the fact that the
Eq.(\ref{harmo}) describes a harmonic function in one dimension as
would be required by the latter solutions when $\xi$ is identified with
the single non-compact space coordinate $x^1$. On the other hand such
identification is consistent with Eq.(\ref{dir1}) which shows that this
non-compact direction is indeed transverse to all branes.

It is instructive to  illustrate by a specific example how the
aforementioned  solutions of the Einstein and field
equations are transformed into themselves by the Weyl transformations
ensuring the uniqueness of $S_{\g_B}$ for conventional and exotic
solutions. We consider the mapping of the Kaluza-Klein momentum in
11-dimensional supergravity, smeared in all transverse directions but
$x^1$ to the exotic smeared membrane with 2 longitudinal times. We
take, as in Table 2, the time to be 9 and the momentum in the direction
8.  The metric is
\begin{equation} ds^2=- \widetilde H^{-1}\, (dx^9)^2+ \widetilde H
\left[ dx^8 +(\widetilde H^{-1}-1)dx^9 \right]^2 +(dx^1)^2
+\sum_{\mu=2\dots7, 10, 11}   (dx^\mu)^2,
  \label{kkmetric}
\end{equation} where $\widetilde H(x^1)$  is
a positive   harmonic function in one dimension
$\widetilde H(x^1)= A+Q |x^1|
$.  In the triangular gauge the relevant vielbein are given by
\begin{eqnarray}
\label{vielkk}e_8{}^{8} &=&  (2-\widetilde H)^{-{1\over 2}}= H^{-{1\over
2}}\, ,\nonumber\\ e_9{}^9 &=& (2-\widetilde H)^{1\over 2}=H^{1\over
2}\, ,\nonumber\\ e_8{}^9 &=&  (2-\widetilde H)^{-{1\over
2}}-(2-\widetilde H)^{1\over 2} =H^{-{1\over 2}}-H^{1\over 2}\, .
\end{eqnarray}
Here we have defined $H=2-\widetilde H$ and the range of $x^1$ is restricted
to $H > 0$. These results, when expressed in terms of $H$, differs from
the vielbein computed in Appendix B1 of
\cite{eh}, where the time direction was 1, by a  sign in the
exponents of the diagonal time and space vielbein. This difference, which is a
consequence of the triangular gauge when the time index is bigger than
the space one, is crucial in Eq.(\ref{weylbrane}) below.
The $p^{\prime a}$ parametrising the Cartan subalgebra  are from
Eq (\ref{vielkk})
\begin{equation}
\label{cartanmkk} p^{\prime 8}=-{1 \over 2} \ln H\, , \qquad   p^{\prime
9}={1
\over 2} \ln H\, ,
\end{equation} while, using Eq.(\ref{vielkk}), one may compute the non-zero
non-Cartan field (see Appendix B1  of
\cite{eh})
\begin{equation}
\label{stepmkk} h_8{}^{\prime 9}= \ln H\, .
\end{equation}
The  vielbein equation Eq.(\ref{vielkk}) also yields
\begin{equation}
\label{vielk}
-[e^{h^\prime}
(de^{-{h^\prime}})]_8^{~9} =d\ln H\, .
\end{equation}
We perform the Weyl reflection $W_{\alpha_{11}}$ using Eq.(\ref{relab}),
(\ref{WeylMe2}) and (\ref{WeylMe1}) . For the Cartan we find  
\begin{eqnarray}
\nonumber
p^a&=&\frac{1}{6}\ln H \qquad a=2,3,4,5,6,7,9\\
\label{weylbrane}
p^a&=&-\frac{1}{3}\ln H \qquad a=8,10,11\, .
\end{eqnarray}
For the non-Cartan fields we find from Eqs.(\ref{relab}), (\ref{vielk}) and
(\ref{weylbrane}), up to a sign,
\begin{equation}
\label{Afield}
dA_{8 \, 10
\, 11} = dH^{-1}.
\end{equation}
Taking into account the relation Eq.(\ref{dir1}) this solution describes
an exotic `membrane' in the directions 8, 10 and 11,  in a phase with
signature $(2,8,-)$ with time directions 10 and 11,  characterised by
the harmonic function
$H=2-\widetilde H$.
Under this Weyl transformation the smeared $KK$-momentum of M-theory is
thus mapped onto an exotic membrane with two longitudinal times of
$M^*$-theory. The exotic membrane obtained is in perfect agreement
with the T-duality interpretation as it can be checked by applying
the Buscher transformations \cite{buscher}
on the KK-momentum solution in 10 dimensions
obtained by reducing  Eq.(\ref{kkmetric}) along the direction 11 and
then uplifting back to  $M^*$.

While Eq.(\ref{weylbrane}) illustrate the invariance of the  quadratic
form of $\g$ restricted to its Cartan subalgebra, 
Eq.(\ref{Afield})  confirms the invariance of the lapse constraint,
yielding
\begin{equation}
\label{branex}
\Lambda= -\frac{1}{2} (d\ln H)^2\, ,
\end{equation}
where, despite the fact that the membrane has two longitudinal times 10
and 11, a minus sign does arise in the quadratic lapse constraint from
the negative kinetic term in the
$(2,8,-)$ signature.

We point out that the Einstein solutions required  the additional
conditions
\begin{equation}
\Theta_E (-1)^{t_A+1}=1\qquad \Theta_E=(-1)^{T+1} \Theta_M\, ,
\label{theta}
\end{equation}
where $\Theta_E$ and $\Theta_M$ are the signs of the kinetic
terms for the electric and magnetic potentials  in the exotic actions.
These conditions are trivially satisfied in the conventional phase
$S_{\g_2}\equiv S_{{\cal G}{}_{(2)}^{++}}^{(1,9,+)}$ where $\Theta_E =1$. This
phase  is characterised by one time which is always longitudinal to the
branes. Therefore they are automatically satisfied in all phases as the
solutions in these phases can be obtained from Weyl transformations and
hence do exist with $\Theta_E$ identified with the corresponding sign
encoded in $\varepsilon$.

\setcounter{equation}{0}
\section{Perspectives}

We have shown that all solutions of the cosmological and the brane
overextended invariant actions $S_{\g_C}$ and $S_{\g_B}$ are solutions
of the very-extended invariant actions $S_{\G}$. The variable $\xi$
used to parametrise the motion of the fields on the coset space
$\G/K^{+++}$ in the non-linear realisation $S_{\G}$ of $\G$ is then
identified respectively to a time  or to a space coordinate and a
set of fields are consistently made to
vanish. This suggests that the generators of $\G$ contain a huge gauge
redundancy. These generators are indeed
associated to fields which must be interpreted as potentials rather
than field strengths and equivalent descriptions may be possible with
both space and time components of gauge fields. The
conventional intersecting extremal branes,  easily described by the
$S_{\g_B}$, could perhaps also be obtained from $S_{\g_C}$ in a
complicated way, while the latter action describes trivially   the
Kasner-like cosmological solutions.

Thus the fact that $\G$ contains on equal footing time and space allows
a selection of the best gauge potentials to describe a
particular solution but this `covariant' approach need not  in
principle comprehend a larger gauge invariant content than a
non-covariant $\g$,  as least as long as only conventional `phases' of
theories are considered. The existence of exotic solutions is a specific
feature of $\G$ which has no counterpart in the overextended action
$S_{\g_C}$ and is a necessary concomitant of $\G$-invariance and
the temporal involution, as suggested in reference \cite{keu1}. The only way
exotic phases could decouple in  very-extended algebra would be by restricting
the signature to be Euclidean, a rather problematic alternative.

An essential result of this work is the existence of solutions of
$S_{\G}$ and $S_{\g_B}$ which are identical to the space-time covariant
solutions of intersecting extremal branes smeared in all directions but
one. This was made possible by the interpretation of  $\xi$ as a
spatial coordinate in the context of $S_{\g_B}$. In this way  we do not
have to restrict the interpretation of  $S_{\G}$ to  algebraic
considerations as in reference \cite{eh}. The reason of the
importance of this result is that the exact solutions of the space-time
covariant theories exist with the same algebraic structure but
different, although simple, functional dependance in more uncompactified
dimensions. Thus we have a laboratory to check whether or not the
Kac-Moody theories can reach, at least in these simple cases,
uncompactified gravity from the information contained in higher levels.
If this turns out not to be the case, then the content of these
approaches would probably only be a consistent dimensional reduction of
the covariant space time theories down to one dimension through an
infinity of equivalent fields. If uncompactified theories are reached,
even at this elementary level, one has attained the first step of a
formulation of gravity coupled to some matter fields, which is
conceptually completely different from the Einstein approach, and which
possibly includes new degrees of freedom which are hinted at by the
string perturbative approaches. We hope to be able to provide the
answer to this question in future work.

\section*{Acknowledgments}

We thank Riccardo Argurio, Hermann Nicolai and Philippe Spindel  for
fruitful discussions. 

This work was supported in part  by the NATO grant PST.CLG.979008, by
the ``Interuniversity Attraction Poles Programme -- Belgian Science
Policy '', by IISN-Belgium (convention 4.4505.86),   by Proyectos
FONDECYT 1020629, 1020832 and 7020832 (Chile) and by the   European
Commission FP6 programme MRTN-CT-2004-005104, in which F.~E., M.~H. and
L.~H. are associated to the V.U.Brussel (Belgium).


\begin{thebibliography}{99}



\bibitem{cjlp} E.~Cremmer, B.~Julia, H.~L{\"u} and C.~N.~Pope, {\it
Higher dimensional origin of D=3 coset symmetries}, {\tt
hep-th/9909099}.


\bibitem{ogw} M. Gaberdiel, D. Olive and P. West, {\it A class of
Lorentzian Kac-Moody algebras},
 Nucl. Phys. {\bf B645} (2002) 403, {\tt hep-th/0205068}.


\bibitem{west01} P. West, {\it $E_{11}$ and M theory}, Class. Quant.
Grav. {\bf 18} (2001) 4443, {\tt hep-th/0104081}.

\bibitem{lw} N.~D.~Lambert and P.~C.~West, {\it Coset symmetries in
dimensionally reduced bosonic string theory}, Nucl. Phys. {\bf B615}
(2001) 117, {\tt hep-th/0107209}.

\bibitem{ehtw} F. Englert, L.~Houart, A.~Taormina and P.~West, {\it
The symmetry of M-theories},
  J. High Energy Phys. {\bf 09} (2003) 020, {\tt hep-th/0304206}.


\bibitem{damourhn00} T. Damour, M. Henneaux and H. Nicolai, {\it
Cosmological billiards}, Class. and Quant. Grav. {\bf 20} (2003) R145
, {\tt   hep-th/0212256}, and references therein.


\bibitem{damourh00}   T. Damour, M. Henneaux, {\it E(10), BE(10) and
arithmetical chaos in   superstring cosmology}, Phys. Rev.Lett. {\bf
86} (2001) 4749, {\tt   hep-th/0012172}; T. Damour, M. Henneaux, B.
Julia and H. Nicolai,   {\it  Hyperbolic Kac-Moody algebras and chaos
in Kaluza-Klein models}, Phys.   Lett. {\bf B509} (2001) 323,  {\tt
hep-th/0103094}.


\bibitem{damourbhs02} T. Damour, S. de Buyl, M. Henneaux and C.
Schomblond,  {\it Einstein billiards and overextensions of
finite-dimensional simple Lie algebras}, J. High Energy Phys. {\bf 08}
(2002) 030, {\tt hep-th/0206125}.


\bibitem{damourhn02} T. Damour, M. Henneaux and H. Nicolai, {\it
$E_{10}$ and a small tension expansion of M-theory}, Phys. Rev. Lett.
{\bf 89} (2002) 221601, {\tt hep-th/0207267}.

\bibitem{eh} F. Englert and L.~Houart, {\it $\G$ invariant formulation
of gravity and M-theories:exact BPS solutions}, J. High Energy Phys. {\bf 01}
(2004) 002, {\tt hep-th/0311255}.



\bibitem{damourn04} T. Damour and H. Nicolai,
{\it Eleven dimensional supergravity and the E10/KE10
sigma-model at low A9 levels}, {\tt   hep-th/0410245}.

\bibitem{nike} A. Kleinschmidt and H. Nicolai,
{\it IIB supergravity and E10},
{\tt   hep-th/0411225}.

\bibitem{west02} P. West, {\it Very Extended E8 and A8 at low levels,
gravity and supergravity}, Class. Quant. Grav. {\bf 20} (2003) 2393,
{\tt hep-th/0212291}.

\bibitem{nifi} H. Nicolai and T. Fischbacher, {\it Low level representations of
E10 and E11}, in: {\it Kac-Moody Lie algebra and related topics},
eds. N. Sthanumoorthy and K.C. Misra, Contempary Mathematics 343, American
Mathematical Society, 2004, {\tt hep-th/0301017}.

\bibitem{weke} A. Kleinschmidt, I. Schnakenburg and P. West,
{\it Very-extended Kac-Moody algebras and their interpretation at low levels},
 Class. Quant. Grav. {\bf 21} (2004) 2493, {\tt hep-th/0309198}.


\bibitem{keu1} A. Keurentjes, {\it $E_{11}$: Sign of the times},
Nucl. Phys. {\bf 697} (2004) 302,  {\tt hep-th/0402090}.

\bibitem{keu2} A. Keurentjes, {\it Time-like T-duality algebra},
J. High. Energy Phys. {\bf 11} (2004) 34, {\tt hep-th/0404174}.

 \bibitem{eh2} F.~Englert and L.~Houart, {\it $\G$ invariant formulation
of gravity and M-theories:exact intersecting brane
solutions}, J. High Energy Phys. {\bf 05} (2004) 059, {\tt hep-th/0405082}.

\bibitem{aeh} R.~Argurio, F. Englert and  L.~Houart, {\it Intersection
rules for p-branes},
 Phys. Lett. {\bf B398} (1997) 61, {\tt hep-th/9701042}.


\bibitem{ah} R.~Argurio and L.~Houart, {\it On the exotic phases of M-theory},
Phys. Lett. {\bf B450} (1999) 109, {\tt hep-th/9810180}


\bibitem{hull1} C.~M.~Hull, {\it Timelike T-duality, de Sitter space,
large N gauge theories and topological field theory},
J. High Energy Phys {\bf 07} (1998) 021, {\tt hep-th/9806146}; C.~M.~Hull and
R.~R.~Khuri, {\it Branes, times and dualities},
Nucl. Phys. {\bf B536} (1998) 219, {\tt hep-th/9808069}.


\bibitem{hull2} C.~M~Hull,
{\it Duality and the signature of space-time}, J. High Energy Phys.
{\bf 11} (1998) 017, {\tt hep-th/9807127}.

\bibitem{exop} C.~M.~Hull and B.~Julia,
{\it Duality and moduli spaces for timelike reductions},
Nucl. Phys. {\bf B534} (1998) 250, {\tt hep-th/9803239};
E.~Cremmer, I.~V.~Lavrinenko, H.~L\"u, C.~N.~Pope, K.~S.~Stelle
and T.~A.~Tran,
{\it Euclidean signature supergravities, dualities and instantons},
Nucl. Phys. {\bf B534} (1998) 40,
{\tt hep-th/9803259}.

\bibitem{weylt} S. Elizur, A. Giveon, D. Kutasov and E. Rabinovici,
{\it Algebraic aspects of matrix theory on $T^d$}, Nucl. Phys. {\bf B509}
(1998) 122, {\tt hep-th/9707217}; N. A. Obers and P. Pioline,
{\it U-duality and M-theory},
Phys. Rept. {\bf 318} (1999) 113-225
{\tt hep-th/9812139}.

\bibitem{banks} T. Banks, W. Fischler and L. Motl,
{\it Dualities versus singularities}, J. High Energy Phys. {\bf 01} (1999)
 019, {\tt hep-th/9811194}.

\bibitem{buscher} T. H. Buscher, {\it A symmetry of the string background
field equations}, Phys. Lett. {\bf B194} (1987) 59; {\it Path integral
derivation of quantum duality in nonlinear sigma models}, Phys. Lett.
{\bf B201} (1988) 466.

\end{thebibliography}
\end{document}